\documentclass[aps,prd,twocolumn,showpacs,showkeys,superscriptaddress,nofootinbib, 10pt]{revtex4-2}

\usepackage[utf8]{inputenc} % Input encoding
\usepackage{amsmath, amssymb} % Math symbols and environments
\usepackage{mathrsfs}
\usepackage{graphicx, subcaption} % Figures and subfigures
\usepackage{hyperref} % Hyperlinks
\usepackage{float} % Float placement
\usepackage{color, caption} % Caption customization
\usepackage{natbib} % Bibliography support
\usepackage{lineno}
\DeclareMathOperator{\Tr}{\rm Tr}
\begin{document}

\title{Gravity as emergent phenomena for spherically symmetric black hole accretion of multi-component flow with relativistic equation of state}

\author{Tuhin Paul}
\email{tuhin.paul96@gmail.com}
\affiliation{Indian Statistical Institute, Kolkata - 700108, India}

\author{Aishee Chakraborty}
\email{aishee.chakraborty@edu.uni.lodz.pl}
\affiliation{University of Lodz, ul. Matejki Jana 21/23, Lodz 90-237, Poland}
\affiliation{Harish-Chandra Research Institute, HBNI, Chhatnag Road, Jhunsi, Allahabad 211 109, India}
\author{Souvik Ghose}
\email{dr.souvikghose@gmail.com}
\affiliation{Harish-Chandra Research Institute, HBNI, Chhatnag Road, Jhunsi, Allahabad 211 109, India}

\author{Tapas K. Das}
\email{das@hri.res.in}
\affiliation{Harish-Chandra Research Institute, HBNI, Chhatnag Road, Jhunsi, Allahabad 211 109, India}

\date{\today}

\begin{abstract}\noindent
We investigate analogue gravity phenomena arising as a result of the linear 
perturbation of the spherically symmetric accretion flows onto 
non rotating black holes, where the gravitational field is determined by a set of 
post Newtonian pseudo Schwarzschild black hole potentials and the infalling matter is described by a relativistic multi-species equation of state. 
The stationary transonic integral accretion solutions corresponding to the steady state of aforementioned type of accreting systems 
are constructed and the stability analysis of such solutions are performed through the time dependent linear perturbation 
of the accretion flow. Such linear stability analysis leads to the formation of a black hole like sonic metric embedded 
within the infalling matter. The acoustic horizons are then identified by constructing the causal structure, i.e., the 
Carter-Penrose diagrams. The variation of the analogue surface gravity corresponding to the aforementioned sonic metric has been studied 
as a function of various parameters governing the accretion flow. 
\end{abstract}

\maketitle

% PACS numbers and keywords

\maketitle

% Main content

\section{Introduction}
Linear perturbation of transonic fluid flow leads to the emergence of a black hole like space time metric inside the fluid. Such metric is perceivable to the perturbation only, and not to the embedding fluid. Study of such  emergent metric, commonly known as the sonic, or the acoustic metric, may help to experimentally realize some of the kinematic features of the general theory of relativity within the laboratory framework in terrestrial set up. Such phenomena is known as analogue gravity \cite{moncrief1980stability, unruh1981experimental, visser1998acoustic, visser_review}. Even for a Newtonian set up (Minkowskian background), the black hole like acoustic metric is found to mimic the properties of curved space time, and gravity may be manifested as an emergent phenomena in the aforementioned analogue space time, which may allow one to study the horizon related physics within the framework of such emergent gravity.

Beyond the terrestrial set up, an analogue model of gravity can be studied for accreting black hole systems as well, where the background spacetime metric is itself curved.  Accretion (see, e.g., \cite{frank2002accretion, kato2008black} for details about the accretion processes in astrophysics) onto black holes are predominantly transonic \cite{liang1980transonic}, and hence the linear perturbation of the mass accretion rate, or certain accretion variables, can generate an analogue space time embedded within the accreting fluid (see, e.g., \cite{das2004analogue,TapasAbraham:2005ah, Cadoni:2005nh, Das:2006an, Mach:2009fd, maity2022carter, Fernandes:2021gkf}, and references therein). Study of emergent gravity (in the present paper, we shall use the phrases `analogue gravity' and `emergent gravity' synonymously) phenomena for accreting black holes is particularly important, since for such analogue systems, both type of horizons, the usual gravitational horizons corresponding to astrophysical black holes, and the acoustic horizons generated by the perturbation of the accreting matter, are present simultaneously.

As of now, all the works present in the literature dealing with the analogue models for accreting black holes, consider the accretion flow to be adiabatic (polytropic) or isothermal, and assume the value of the adiabatic index $ \gamma$ (the ratio of the specific heats at constant pressure and constant volume, respectively, i.e., $\gamma=c_p/c_v$) for the flow to be constant in space, i.e., no variation of $\gamma$ with the radial distance (as measured from the black hole event horizon along the flow lines) has been considered in aforementioned works. Also the usual equation of state (EoS hereafter) for polytropic flow, i.e., $p=K{{\rho}^\gamma}$, where $p$ is the fluid pressure, $\rho$ is the density, and $K$ is a constant which is a measure of the entropy of the flow (\cite{pitaevskii2012physical}), does not fully take care of the relativistic state of matter. In addition, accreting matter may consists of
various species, like electron, ion, and positron. Considering all such limitations of the equation of state used to describe the accretion flow, it is tempting to introduce an equation of state which will be efficient to describe the thermodynamic properties of relativistically flowing fluid (dynamics of matter at the vicinity of compact objects, being it accreting fluid or astrophysical jets, travel with very large velocity, at a velocity comparable to the velocity of light in vacuum), as well as will be able to take care of the radial variation of the heat capacities (and thus of the corresponding polytropic index) of multi component matter. Several works addressed many of such issues (\cite{chandra, synge, taub, mathews1971hydromagnetic, mignone, ryu2006, ic2009effects}) where attempts were made to propose a relativistic equation of state with variable $\gamma$, some of them also considering a multi-component fluid. Among such models, the equation of state proposed by Ryu and his collaborators \cite{ryu2006, ic2009effects, kumar2013effect} seems to have a relatively simple form (by maintaining all the necessary elegance to handle a relativistic flow with variable $\gamma$) suitable to study transonic accretion onto compact objects.  

In our present work, we shall construct  the stationary transonic integral accretion solutions, and draw the corresponding phase portraits  for spherically symmetric flow of ideal fluid onto non rotating astrophysical black holes, where the accretion dynamics is governed by all four post Newtonian pseudo Schwarzschild black hole potentials (as proposed by \cite{pw1980, artemova1996, nowak1991}) available in the literature (see, e.g., \cite{das2002generalized}for a detailed description of all such potentials), and the thermodynamics of the flow is controlled by the equation of state as described by Ryu and his collaborators (\cite{ryu2006, ic2009effects}). We then perturb the aforementioned solutions to verify whether the assumed steady state remains stable subjected to the introduction  of  radial perturbation. We observe that the perturbation propagates in the form of the propagation of  a massless scalar field in a curved pseudo-Riemannian background, and such propagation is governed by a black hole-like acoustic metric which possesses an acoustic horizon.  Causal structures (the Carter-Penrose (\cite{wald1984general, cp_visser}) diagrams) are constructed to identify the sonic horizons and to study the horizon related properties corresponding to the sonic metric. In this way we study the spherically symmetric multi-component black hole accretion as a classical analogue gravity model for relativistic equation of state (with a more realistic description of $\gamma$), which has not been done before in the literature.    

In the next section, following \cite{ryu2006, ic2009effects, kumar2013effect}, we describe the basic features of the relativistic equation of state used in our work. In subsequent sections, we describe the nature of the black hole potentials used in the work, the equations governing the dynamics of the accretion flow, and develop the solution scheme which can be used to realize the complete nature of the transonic properties of the stationary integral accretion solutions. We then provide the detailed description of our stability analysis scheme for the flow, and how the emergent gravity phenomena can be conceived from such analysis. Lastly, we define the acoustic surface gravity corresponding to the analogue metric, and will study how the value of such quantity gets influenced by the conserved specific energy of the flow ${\cal E}$, as well as with the variable (space variation) flow temperature (and thus with the distance dependent adiabatic index of the flow). At last we conclude by providing a tentative future direction for our forthcoming works. 
\section{On Multi-species Equation of State}
\label{sec:meos}
In this section, we summarize various properties of the EoS used in our work. We consider a multi-species accretion flow consisting of electrons $(e^{-} )$, positrons $(e^{+} )$ and protons $(p^{+} )$ of proportion parameterized by $\xi$. The number density ($n$) of the accreting matter is given by,
\begin{equation}
    \label{numden}
    n=\sum_i n_i=n_{e^-}+n_{e^+}+n_{p^+}
\end{equation}
where $n_{e^-}$ , $n_{e^+}$ and $n_{p^+}$ are the electron, positron and proton number densities, respectively.
If we further demand that the overall charge-neutrality is always maintained, we have:
\begin{equation}
    \label{chgn1}
    n_{e^-}=n_{e^+}+n_{p^+}
\end{equation}
which gives:
\begin{equation}
    \label{chgn2}
    n=2n_{e^-}, \;\;\; n_{e^+}=n_{e^-}(1-\xi),
\end{equation}
where,  $\xi  \equiv \frac{n_{p^+}}{n_{e^-}} $.\\
The mass density is expressed as, 
\begin{equation}
    \label{rho}
    \rho=\sum_i n_i m_i = n_{e^-}m_{e^-} k = \rho_{e^-}k,
\end{equation}
where, $\eta=\frac{m_{e^-}}{m_{p^+}}$, $m_{e^-}$, $m_{p^+}$ are the electron and proton masses, respectively and $k=2-\xi(1-\frac{1}{\eta})$.
\newline
The EoS for multi-species flow as given in \cite{ic2009effects}:
\begin{equation}
    \label{energyav}
    \bar{e}=\sum_i e_i= \sum_i n_im_ic^2+p_i \left(\frac{9p_i+3n_im_ic^2}{3p_i+2n_im_ic^2}\right)
\end{equation}

where, $\bar{e}$ is the energy density.

The isotropic pressure can be defined with respect to the Boltzmann constant $K_B$ and absolute temperature $T$ as,
\begin{equation}
    \label{pressure}
    p=\sum_i p_i=2 n_{e^-} K_BT
\end{equation}
We define the dimensionless temperature as:
\begin{equation}
    \label{theta}
    \Theta=\frac{K_BT}{m_{e^-}c^2}
\end{equation}
Using eq. (\ref{theta}), the expression for the isotropic pressure takes the form: 
\begin{equation}
    \label{pressure2}
    p=2 n_{e^-}m_{e^-} c^2\Theta=2 \rho_{e^-}c^2\Theta = 2 \frac{\rho}{k}   c^2\Theta
\end{equation}
The EoS given in eq. (\ref{energyav}) can be written as:
\begin{equation}
    \label{energyav2}
    \Bar{e}=n_{e^-}m_{e^-}c^2f=\rho_{e^-}c^2f=\frac{\rho}{k} c^2f
\end{equation}
where, $f=(2-\xi)\left[1+\Theta\left(\frac{9\Theta+3}{3\Theta+2}\right)\right]+\xi \left[ \frac{1}{\eta}+\Theta \left(  \frac{9\Theta+\frac{3}{\eta}}{3\Theta+\frac{2}{\eta}}\right)\right]$\\
The specific enthalpy can be written as: 
\begin{equation}
    \label{enthalpy}
    h=\frac{\Bar{e}+p}{\rho}=\frac{c^2}{k}(f+2\Theta)
\end{equation}
The polytropic index is conveniently given by \cite{ic2009effects}:
\begin{equation}
    \label{polytrop}
    N=\frac{1}{2}\frac{df}{d\Theta}
\end{equation}
Similarly, the adiabatic Index can be written as:
\begin{equation}
    \label{adb}
    \Gamma=1+\frac{1}{N}
\end{equation}
The adiabatic sound speed is then given by:
\begin{equation}
    \label{soundspeed}
    c_{\rm s}^2=\frac{2\Theta \Gamma} k
\end{equation}

\begin{figure*}
    \centering
    \includegraphics[width=0.5\linewidth]{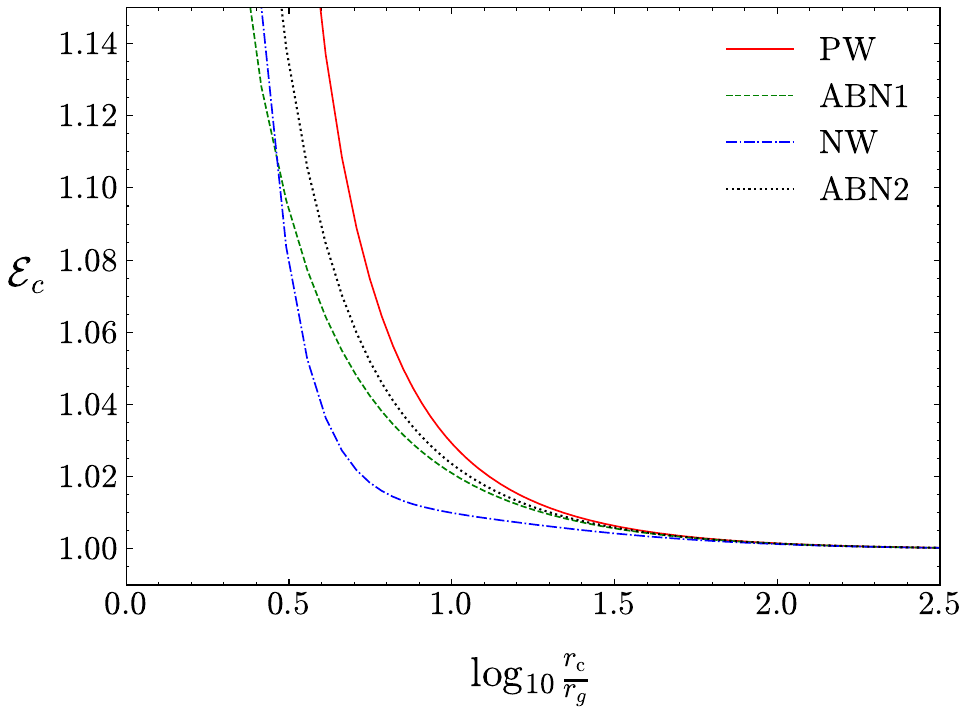}
    \caption{Variation of critical points with energy at the critical point ($\mathcal{E}_c$). In the present case, the energy is conserved, i.e., $\mathcal{E}_c = \mathcal{E}$. Radial distace is measured in the unit of $r_g=\frac{2GM}{c^2}$. Refer to eqs (\ref{eq:pw})- (\ref{eq:abn2}) for the descriptions of the potentials.}
    \label{fig:rad_en}
\end{figure*}

\section{Stationary Solution of the flow}
In this section, we are going to obtain the stationary integral solutions of the corresponding steady state spherical accretion flow around a Schwarzschild black hole. Such stationary solutions will be obtained considering four different gravitational potentials commonly known as modified Newtonian potentials in the literature:

\begin{align}
    \label{eq:pw}
    \phi_{\rm PW} &= -\frac{1}{2(r-1)} \ \\  
    \label{eq:abn1}
    \phi_{\rm ABN1}&= -1 + \left(1-\frac{1}{r}\right)^{\frac{1}{2}}  \\ 
    \label{eq:nw}
    \phi_{\rm NW} &= -\frac{1}{2r} \left[1-\frac{3}{2r} + 12\left(\frac{1}{2r}\right)^2 \right]\\  
    \label{eq:abn2}
    \phi_{\rm ABN2} &= \frac{1}{2} \ln \left(1-\frac{1}{r}\right)
\end{align}
The first potential in eq. (\ref{eq:pw}) (i.e., $\phi_1$, hereafter PW) was proposed by Paczy\'nski and Wiita \cite{pw1980}. $\phi_2$ and $\phi_4$ (eqs (\ref{eq:abn1}) and (\ref{eq:abn2})) were proposed by Artemova, Bj\.ornsson and Novikov \cite{artemova1996} (hereafter ABN1 and ABN2 respectively). $\phi_3$ (eq. (\ref{eq:nw})) was proposed by Nowak and Wagoner \cite{nowak1991} (hereafter, NW). Here, we have used the units $G=c=M_{bh}=1$, and all the radial distances are scaled by the Schwarzschild gravitational radius $r_g=\frac{2GM_{bh}}{c^2}$, where $M_{bh}$ denotes the mass of the accretor (black hole).
In the following, we will write all the necessary equations governing the fluid flow for a general arbitrary gravitational potential $\phi$. To proceed, we need to evaluate the advective velocity gradient of the accretion flow. To do that, we need two conserved quantities, one is Bernouli's constant or specific energy in our case, and the other one is mass accretion rate. 

The expression for the specific energy of a fluid moving with a  radial velocity $u$ in an arbitrary gravitational potential $\phi(r)$, is:
\begin{equation}
    \mathcal{E} = \frac{u^2}{2} + h + \phi(r).
\label{General sp. energy}
\end{equation}
And, the expression of conserved mass accretion rate for the flow is:
\begin{equation}
    \dot M = 4\pi r^2 \rho(r) u(r)
    \label{mass accretion rate}
\end{equation}

First, by differentiating the expression of specific energy in eq. (\ref{General sp. energy}) we get the following expression:

\begin{equation}
    u \frac{du}{dr} + \frac{dh}{dr} + \frac{d\phi}{dr} = 0
    \label{ad1}
\end{equation}

Secondly by differentiating logarithmic expression of conserved mass accretion rate in eq. (\ref{mass accretion rate}), we get an expression:

\begin{equation}
    \frac{1}{u} \frac{du}{dr} + \frac{1}{\rho} \frac{d\rho}{dr} + \frac{2}{r} = 0
    \label{ad2}
\end{equation}

From equations (\ref{ad1}) and (\ref{ad2}) we can get an expression for advective velocity gradient $\frac{du}{dr}$ of the flow. But till now, the advective enthalpy gradient is unknown. Let's proceed to evaluate that expression.

Define specific energy $e = \frac{\bar e}{\rho}$ and from the first law of thermodynamics we know that,
\begin{equation}
    \label{thermo_1_1}
    \frac{de}{dr} - \frac{p}{\rho^2}  \frac{d\rho}{dr} = 0
\end{equation}
Using eqs (\ref{pressure2}), (\ref{energyav2}), and (\ref{polytrop}) in eq. (\ref{thermo_1_1}), we obtain:,
\begin{equation}
    \label{eq:new}
    \frac{1}{\rho}\frac{d\rho}{dr} = \frac{N}{\Theta}\frac{d\Theta}{dr}
\end{equation}
Using eqs (\ref{pressure2}), (\ref{energyav2}), (\ref{polytrop}), and (\ref{thermo_1_1}) one can write:
\begin{equation}
\label{de_dr}
\frac{de}{dr} = \frac{2N}{k} \frac{d\Theta}{dr}
\end{equation}
Also, from the definition of enthalpy in eq. (\ref{enthalpy}) we can get the advective enthalpy gradient expression:
\begin{equation}
    \frac{dh}{dr} =  \left[ \frac{2(N + 1)}{k} \right] \frac{d\Theta}{dr}
    \label{dh_dr}
\end{equation}
Using eq. (\ref{dh_dr}) in eq. (\ref{ad1}) and eq. (\ref{eq:new}) in eq. (\ref{ad2}) some algebraic manipulation (elimination of $\frac{d\Theta}{dr}$ from both of the equations) leads to the final expression for advective velocity gradient for a general gravitational potential $\phi$:

\begin{equation}
    \frac{du}{dr} = \frac{ \frac{2c_{\rm s}^2}{r} - \frac{d\phi}{dr} }{u - \frac{c_{\rm s}^2}{u}} = \frac{\mathcal{N}}{\mathcal{D}}
    \label{velocity_grad_general}
\end{equation}
The denominator $\mathcal{D}$ of the above expression of advective velocity gradient is zero at critical point, $u = c_{\rm s}$. The radius of such a critical point, which can also be called sonic point in this case is denoted by $r_{\rm c}$ hereafter. This radius depends on the energy of the flow and its variation with energy is shown in fig. (\ref{fig:rad_en}). As the velocity profile is smooth everywhere the first derivative of $u$ should be finite everywhere in the domain. So, where $\mathcal{D} = 0$, the numerator $\mathcal{N}$ has to be zero too, which gives us another critical condition:
\begin{equation}
   \left. c_{\rm s}^2 \right \vert_{r_{\rm c}} = \left.\frac{r}{2} \frac{d\phi}{dr} \right \vert_{r_{\rm c}}=\left. u_{\rm c}^2 \right \vert_{r_{\rm c}}
    \label{crit2}
\end{equation}
By integrating the expression of $\frac{du}{dr}$ numerically, we can obtain the stationary solutions and draw the phase profiles for each of the four above mentioned gravitational potentials. But in order to accomplish this task we also need to consider an expression for $\frac{d\Theta}{dr}$ and integrate it simultaneously with $\frac{du}{dr}$. In the next session we elaborate on the methodology of obtaining such stationary solutions.\\
Similar calculations involving the elimination of $\frac{du}{dr}$ from the eq: (\ref{ad1}) and eq: (\ref{ad2}) leads to the expression of $\frac{d\Theta}{dr}$ in terms of an arbitrary general gravitational potential in the form:
\begin{equation}
     \label{eq:dthetadr_gen}
    \frac{d\Theta}{dr}=\frac{\frac{2\Theta }{N}\left(\frac{u^2}{r}-\frac{1}{2}\frac{d\phi}{dr}\right)}{c_{\rm s}^2-u^2}
\end{equation}
Analytical expressions specific to the different potentials are found in Appendix \ref{appena} and the Mach number vs radius profiles for all the four potentials are given in fig. (\ref{fig:main1}). Dynamical system analysis involving this general gravitational potential is addressed in the Appendix \ref{appenb} .

% \section{Stationary Solution \& Phase Portrait}
% In what follows we demonstrate the methodology of obtaining phase profiles and analyze the nature of the critical point of the flows considering the PW potential only. For any other pseudo-potential the scheme remains the same and the relevant equations are placed in Appendix A. 
The expression for the specific energy for a flow in steady state is given by:
\begin{equation}
\label{eq:euler1}
    \mathcal{E} =\frac{u^2}{2}+h-\frac{1}{2(r-1)}.
\end{equation}
\begin{figure*}[ht]
    \centering
    \begin{subfigure}[b]{0.45\textwidth}
        \centering
        \includegraphics[width=\textwidth]{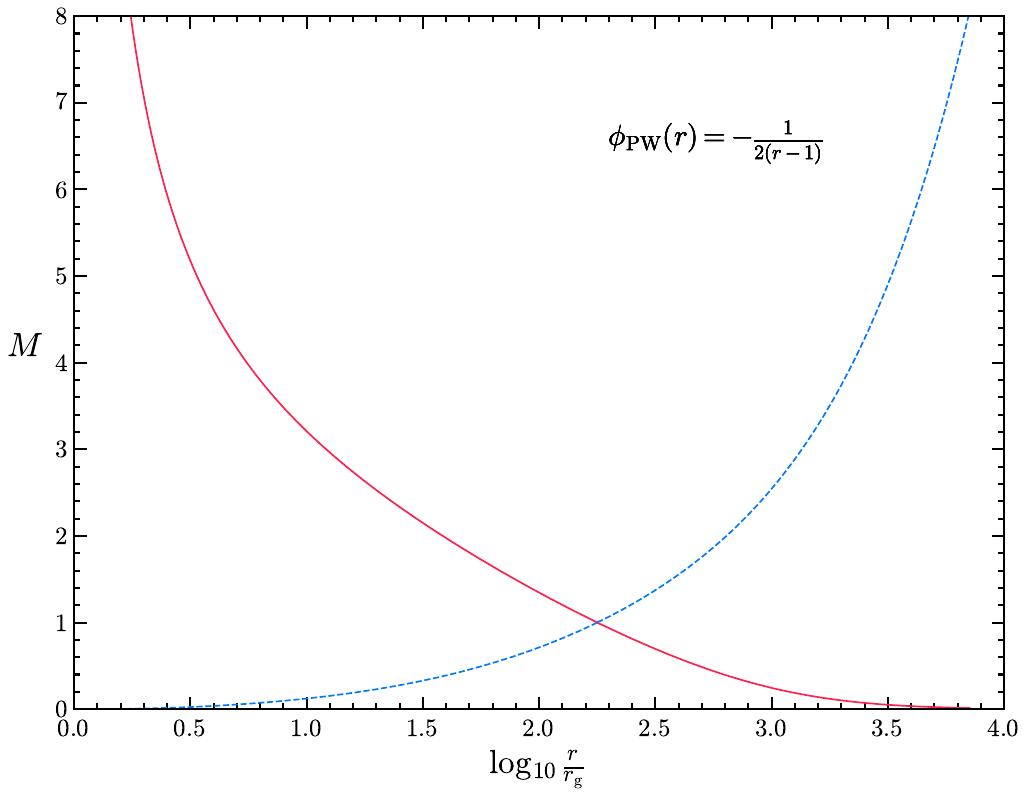}  % Replace with your figure file
        \caption{}
        \label{fig:a}
    \end{subfigure}
    \hfill
    \begin{subfigure}[b]{0.45\textwidth}
        \centering
        \includegraphics[width=\textwidth]{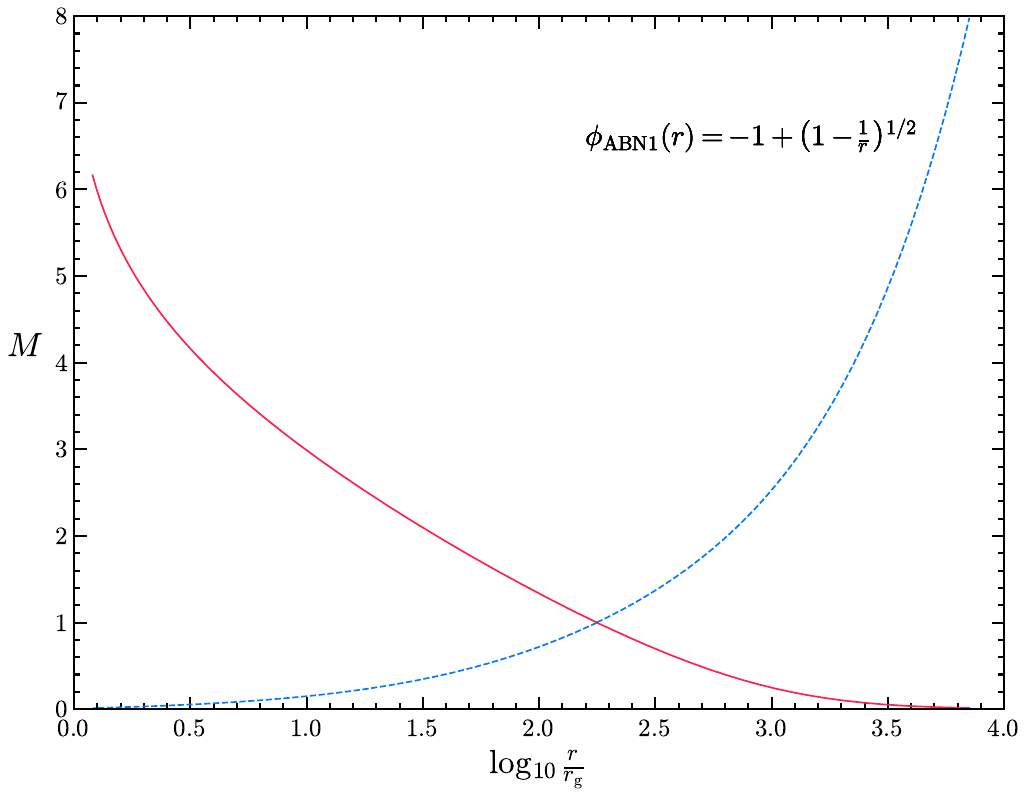}  % Replace with your figure file
        \caption{}
        \label{fig:b}
    \end{subfigure}
    \vskip\baselineskip
    \begin{subfigure}[b]{0.45\textwidth}
        \centering
        \includegraphics[width=\textwidth]{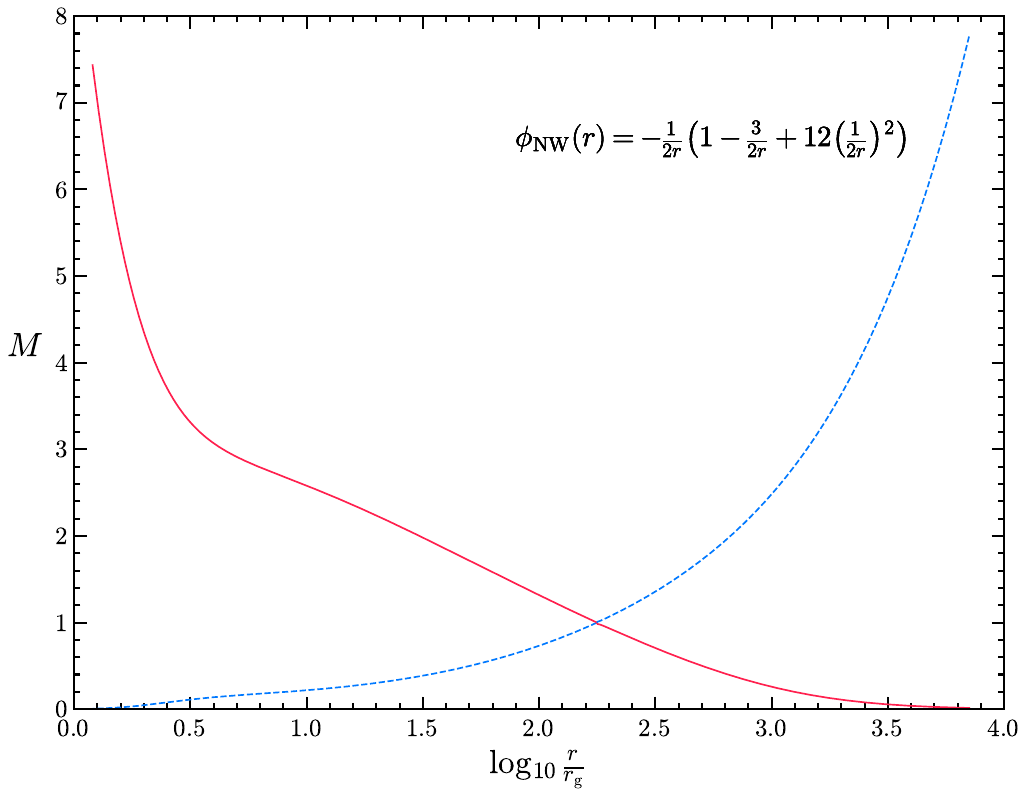}  % Replace with your figure file
        \caption{}
        \label{fig:c}
    \end{subfigure}
    \hfill
    \begin{subfigure}[b]{0.45\textwidth}
        \centering
        \includegraphics[width=\textwidth]{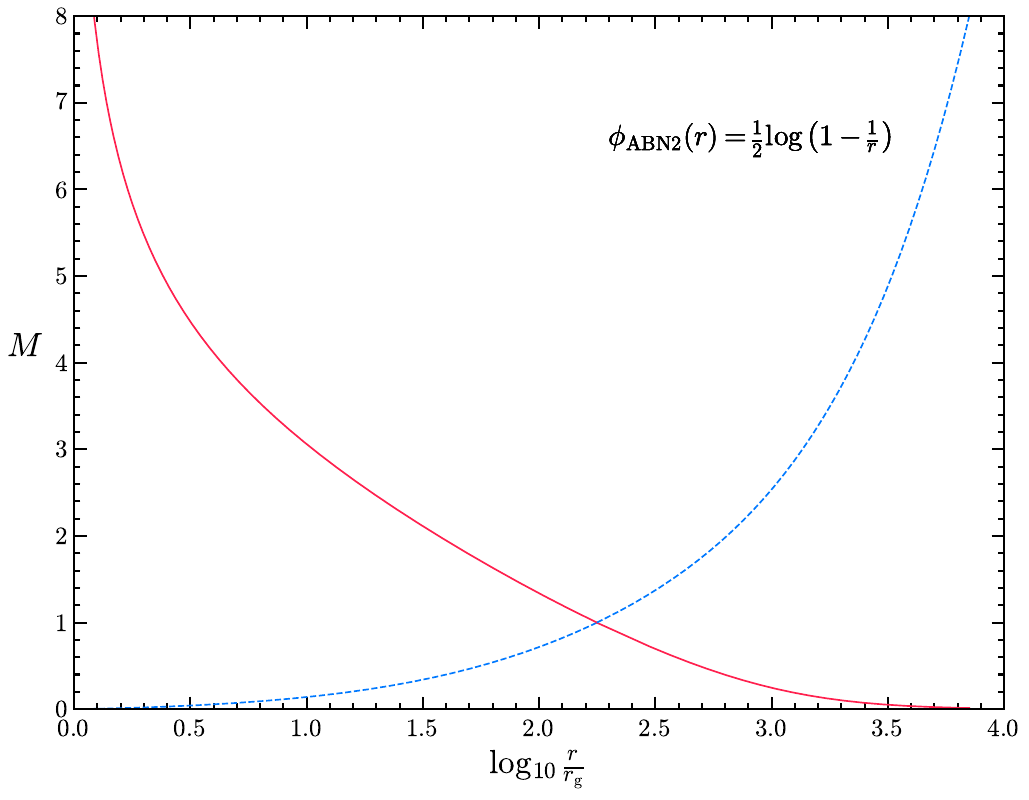}  % Replace with your figure file
        \caption{}
        \label{fig:d}
    \end{subfigure}
    
    \caption{Phase profile for different pseudo-potentials. There is a slight variation of the sonic points (for a demonstrative energy, $\mathcal{E}=1.0007$). Along $y$ -axis we plot the Mach number ($M=u/c_{\rm s}$) of the flow: (a)PW: $r_{\rm c} = 179.57$, (b) ABN1: $r_{\rm c} = 175.905$, (c) NW: $r_{\rm c} = 166.98$, and (d)ABN2: $r_{\rm c} = 177.13$. For all the pseudo-potentials we obtain a single saddle type critical(sonic) point outside the horizon. Radial distance ($r$) is scaled by ($r_g = \frac{2GM}{c^2}$). Solid red lines represent the relevant accretion branch. Refer to eqs (\ref{eq:pw})- (\ref{eq:abn2}) for the descriptions of the potentials.}
    \label{fig:main1}
\end{figure*}
\section{Transonic Accretion from a Dynamical Systems Perspective}
\label{dyn-sec}
In this section we analyze the system for the particular case of $\phi =\phi_1$ (PW potential, eq. \ref{eq:pw}) for demonstration. The general formulation is given in Appendix B. In order to use the analysis standard to Dynamical systems, one can introduce a new variable $\tau$ and write:
\begin{equation}
    \frac{du}{dr}=\frac{du/d\tau}{dr/d\tau}=\frac{\mathcal{N}}{\mathcal{D}}
\end{equation}
Then,
\begin{align}
\label{eq:dyn1}
    \frac{d}{d\tau}(\delta u)&=\frac{\partial \mathcal{N}}{\partial r} \delta r+\frac{\partial \mathcal{N}}{\partial \Theta} \delta \Theta \notag \\ &= \left(-\frac{2c_{\rm s}^2}{r^2}+\frac{1}{(r-1)^3}\right)\delta r 
    +\frac{4c_{\rm s}c_{\rm s}'}{r}\delta \Theta,
\end{align}
where, $c_{\rm s}^{\prime}$ is the partial derivative of $c_{\rm s}$ with respect to $\Theta$:
\begin{equation}
    \label{eq:csprime}
    c_{\rm s}^{\prime}=\frac{1}{2c_{{\rm s}}}\left[\frac{2\Gamma}{k}-\frac{\Theta}{N^2k}\frac{d^2f}{d\Theta^2}\right]
\end{equation}
\begin{align}
\label{eq:dyn2}
     \frac{d}{d\tau}(\delta r) &=\frac{\partial \mathcal{D}}{\partial u} \delta u+\frac{\partial \mathcal{D}}{\partial \Theta} \delta \Theta \notag \\
     &=\left( 1+\frac{c_{\rm s}^2}{u^2}\right)\delta u+\left( - \frac{2c_{\rm s}c_{\rm s}'}{u}\right)\delta \Theta
\end{align}
The expression for the specific energy for a flow in steady state is given by:
\begin{equation}
\label{eq:euler1}
    \mathcal{E} =\frac{u^2}{2}+h-\frac{1}{2(r-1)}.
\end{equation}
The mass accretion rate $\dot{M}$ for steady state flow has the same form as of eq. (\ref{mass accretion rate}). Since $\dot{M}$ is a constant, absorbing the $4\pi$ factor, one can write:
\begin{equation}
\label{eq:massaccr}
    \Dot{M}=\rho u r^2
\end{equation}
Differentiation of eq. (\ref{eq:euler1}) with respect to $r$ gives:
\begin{equation}
\label{eq:eulerdiff}
    u\frac{du}{dr}+\frac{2}{k}(N+1)\frac{d\Theta}{dr}+\frac{1}{2(r-1)^2}=0
\end{equation}
Taking $\log$ at both sides of eq. (\ref{eq:massaccr}) and then taking derivative with respect to $r$, one obtains: 
\begin{equation}
\label{eq:accrdiff}
    \frac{1}{u}\frac{du}{dr}+\frac{N}{\Theta}\frac{d\Theta}{dr}+\frac{2}{r}=0
\end{equation}
Using the value of $\frac{d\Theta}{dr}$ from  eq. (\ref{eq:accrdiff}), eq. (\ref{eq:dyn2}) becomes:
\begin{equation}
\label{delta_theta}
    \delta \Theta=-\frac{\Theta}{Nu}\delta u -\frac{2\Theta}{Nr}\delta r
\end{equation}

\begin{figure*}
    \includegraphics[width = 0.5\linewidth]{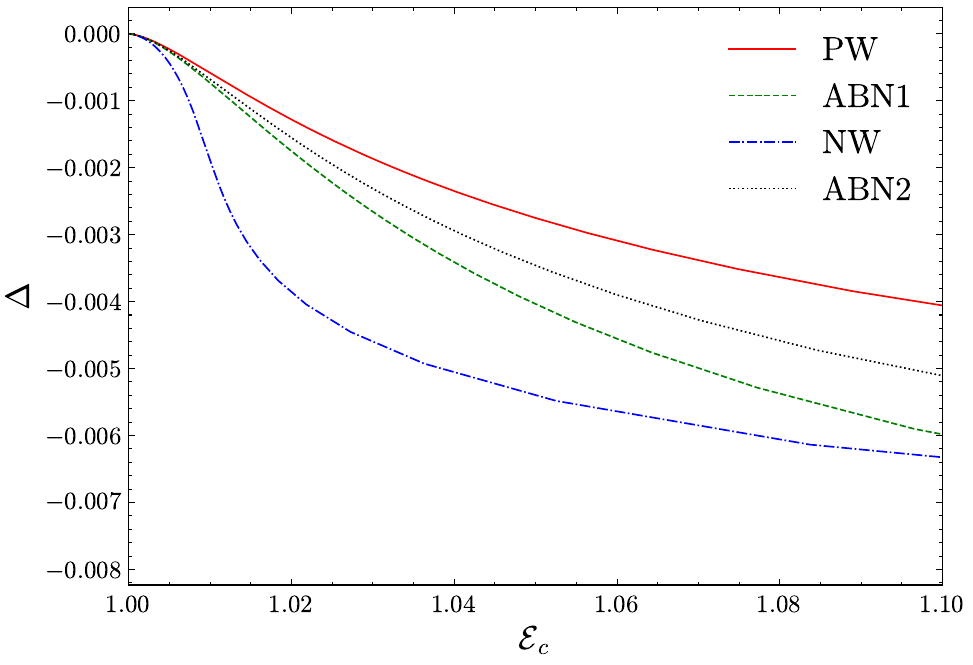}
    \caption{Variation of ${\rm \Delta}$ with $\mathcal{E}_c$ for different pseudo Newtonian potentials. For realistic energy range the critical point in each case remains saddle type in nature (${\rm \Delta} < 0$). Refer to eqs (\ref{eq:pw})- (\ref{eq:abn2}) for the descriptions of the potentials.}
    \label{fig:main}
\end{figure*}

Substituting this value in eq. (\ref{eq:dyn1}) and eq. (\ref{eq:dyn2}) we have,
\begin{align}
    \frac{d}{d\tau}(\delta u)&=\left( -\frac{4c_{\rm s}c_{\rm s}'\Theta}{Nur}\right) \delta u \notag \\
    &+ \left(-\frac{2c_{\rm s}^2}{r^2}+\frac{1}{(r-1)^3}-\frac{8c_{\rm s}c_{\rm s}'\Theta}{Nr^2}\right)\delta r
\end{align}
\begin{equation}
     \frac{d}{d\tau}(\delta r)=\left( 1+\frac{c_{\rm s}^2}{u^2}+\frac{2c_{\rm s}c_{\rm s}'\Theta}{Nu^2}\right)\delta u+\frac{4c_{\rm s}c_{\rm s}'\Theta}{Nur}\delta r
\end{equation}
\begin{align}
    \begin{pmatrix}
         \frac{d(\delta u)}{d \tau} \\
         \frac{d(\delta r)}{d \tau}
     \end{pmatrix} &=
     \begin{pmatrix}
        - \frac{4c_{\rm s}c_{\rm s}'\Theta}{Nur} & -\frac{2c_{\rm s}^2}{r^2}+\frac{1}{(r-1)^3}-\frac{8c_{\rm s}c_{\rm s}'\theta}{Nr^2} \\
         1+\frac{c_{\rm s}^2}{u^2}+\frac{2c_{\rm s}c_{\rm s}'\Theta}{Nu^2} & \frac{4c_{\rm s}c_{\rm s}'\Theta}{Nur} 
     \end{pmatrix}\notag \\ & \times\begin{pmatrix}
             \delta u \\
             \delta r
         \end{pmatrix}
\end{align}
Compactly this can be written as $X'=AX$. If we assume solutions of this dynamical system of equations as $\delta u \sim \exp{\Omega \tau} $ and $\delta r\sim \exp{\Omega \tau} $  we obtain an
eigenvalue equation for the above as $AX = \Omega X$, and the eigenvalue are  obtained from the characteristic equation
$\det (A - \Omega I) = 0$. This gives a quadratic equation
\begin{equation}
    \label{eq:om_quad}
    \Omega^2+ \Tr(A)\Omega+\Delta=0
\end{equation}
where $\Delta = \det (A)$. Since, all the components of $A$ are functions of various flow variables as defined at critical points, all
are completely specified by the flow constants. Hence eq. (\ref{eq:om_quad})  yield the nature of the critical points
themselves. Clearly, the nature of the roots of eq. (\ref{eq:om_quad}) depend on the numerical values of $\Delta$ and the discriminant
$D = ({\rm Tr} A)^{2}-  4{\rm \Delta}$ .\\ For the present case, if we evaluate the matrix elements at the critical points and calculate the values of ${\rm Tr}(A)$ and ${\rm \Delta}$, we observe that
\begin{equation}
\begin{split}
\left. {\rm \Tr}(A)\right \vert_{u=u_{\rm c}, \; c_{\rm s}=c_{\rm sc}}=0 ~~~~~~~~~~~~~~~\\
\left .{\rm \Delta} \right \vert_{u=u_{\rm c}, \; c_{\rm s}=c_{\rm sc}}=-\frac{r_{\rm c}+1}{r_{\rm c}(r_{\rm c}-1)^3}+\frac{4c_{\rm sc}c_{\rm sc}'\Theta_{\rm c}}{N_{\rm c}}\frac{(3r_{\rm c}-5)}{r_{\rm c}^2(r_{\rm c}-1)}
\end{split}
\end{equation}
Thus as long as $\Delta < 0$, we get two real roots of opposite sign, ensuring a saddle type critical point. The variation of $\Delta$ with critical energy ($\mathcal{E}_c$) for all the four potentials  is shown in fig. (\ref{fig:main}). For a reasonable range of $\mathcal{E}_c$, the critical points are always of the saddle type.
% \begin{figure*}[ht]
%     \centering
%     \begin{subfigure}[b]{0.45\textwidth}
%         \centering
%         \includegraphics[width=\textwidth]{f21.pdf}  % Replace with your figure file
%         \caption{}
%         \label{fig:a1}
%     \end{subfigure}
%     \hfill
%     \begin{subfigure}[b]{0.45\textwidth}
%         \centering
%         \includegraphics[width=\textwidth]{f22.pdf}  % Replace with your figure file
%         \caption{}
%         \label{fig:b1}
%     \end{subfigure}
%     \vskip\baselineskip
%     \begin{subfigure}[b]{0.45\textwidth}
%         \centering
%         \includegraphics[width=\textwidth]{f23.pdf}  % Replace with your figure file
%         \caption{}
%         \label{fig:c1}
%     \end{subfigure}
%     \hfill
%     \begin{subfigure}[b]{0.45\textwidth}
%         \centering
%         \includegraphics[width=\textwidth]{f24.pdf}  % Replace with your figure file
%         \caption{}
%         \label{fig:d1}
%     \end{subfigure}
    
%     \caption{Variation of ${\rm \Delta}$ with $\mathcal{E}_c$ for different pseudo Newtonian potentials. (a) PW, (b) ABN1, (c) NW, and (d) ABN2. For realistic energy range the critical point in each case remains saddle type in nature (${\rm \Delta} < 0$). Refer to eqs (\ref{eq:pw})- (\ref{eq:abn2}) for the descriptions of the potentials.}
%     \label{fig:main}
% \end{figure*}

\section{Stability of the stationary solutions}
In order to study the stability of the steady state solutions we look into the evolution of the time dependent perturbations of the equations governing the fluid flow. As long as the perturbations do
not diverge, the steady state solutions remain meaningful.\\ The general form of the continuity equation is:
\begin{equation}
    \frac{\partial \rho}{\partial t}+\nabla .(\rho \mathbf{u})=0
    \label{conti_gen}
    \end{equation}
    The Euler equation, for irrotational flow $(\nabla \times \mathbf{u}=0)$, takes the form :
    \begin{equation}
        \frac{\partial u }{\partial t}+\frac{1}{2}\nabla(u^2)+\nabla h +\nabla(\phi)=0
    \end{equation}
    $\phi$ being the potential of the body force, e.g. gravity. Since the flow is irrotational, one can introduce a velocity potential $\psi$, such that $\mathbf{u}=-\nabla \psi$. \\
Then the Euler equation has the expression (upon integration):
\begin{equation}
     -\frac{\partial \psi }{\partial t}+\frac{1}{2}(\nabla \psi)^2+h +\phi=0
     \label{euler_gen}
\end{equation}
This is a version of Bernoulli's equation. Note that any arbitrary function of time that may come from the integration of the Euler's equation can be absorbed within the velocity potential $\psi$ since it is not unique \cite{visser_review,landau1987fluid}.Considering some arbitrary but exact stationary solutions of the equations of motion $[\rho_{\rm 0}, u_{\rm 0}, \psi_{\rm 0}]$, One can then study the linearized perturbations around this background solution by writing:
\begin{equation}
\begin{split}
    \rho(r,t)= \rho_{\rm 0}(r)+\epsilon \Tilde{\rho}(r,t) \\
    p(r,t)= p_{\rm 0}(r)+\epsilon \Tilde{p}(r,t)\\
     \psi(r,t)= \psi_{\rm 0}(r)+\epsilon \Tilde{\psi}(r,t)\\
     \mathbf{u(r,t)} = \mathbf{u_{\rm 0}}(r)+\epsilon \mathbf{\Tilde{u}}(r,t)
     \end{split}
\end{equation}

where the overhead tilde denotes the perturbed quantity and $0 < \epsilon \ll 1$. Whereas, quantities with subscript $0$ correspond to stationary (background) values. Substituting these into the equations of motion (eq. \ref{conti_gen} and eq. \ref{euler_gen} ) we can write down the equation that governs the propagation of $\Tilde{\psi}(r,t)$:
%\begin{align}
%&\frac{\partial}{\partial t}\left[\frac{\rho_{\rm 0}}{c_{\rm s0}^2}\left( \frac{\partial \Tilde{\psi}(r,t) }{\partial t}+\mathbf{u_{\rm 0}}.\nabla \Tilde{\psi}(r,t) \right) \right]  \notag \\
%& +\nabla\left( \frac{\rho_{\rm 0}}{c_{\rm s0}^2} \left( \frac{\partial \Tilde{\psi}(r,t) }{\partial t}+\mathbf{u_{\rm 0}}.\nabla \Tilde{\psi}(r,t) \right) \mathbf{u_{\rm 0}}-\rho_{\rm 0} \nabla\Tilde{\psi}(r,t) \right)=0    
%\end{align}
%This can be rearranged and simplified as:
\begin{align}
     &\frac{\partial}{\partial t}\left (\frac{\rho_{\rm 0}}{c_{\rm s0}^2} \frac{\partial \Tilde{\psi}(r,t) }{\partial t}\right)+ \frac{\partial}{\partial t}\left ( \frac{\rho_{\rm 0}}{c_{\rm s0}^2} \mathbf{u_{\rm 0}}.\nabla \Tilde{\psi}(r,t)\right) \notag \\
     &+\nabla.\left( \frac{\rho_{\rm 0}}{c_{\rm s0}^2} \mathbf{u_{\rm 0}} \frac{\partial \Tilde{\psi}(r,t) }{\partial t}\right)\notag \\
     &+\nabla.\left(\frac{\rho_{\rm 0}}{c_{\rm s0}^2} \mathbf{u_{\rm 0}}(\mathbf{u_{\rm 0}}.\nabla\Tilde{\psi}(r,t))\right) -\nabla(\rho_{\rm 0} \nabla \Tilde{\psi}(r,t))=0
\end{align}
In a more compact and precise form,
\begin{equation}
\label{eq:pertub}
    \partial_\mu(f^{\mu \nu}\partial_\nu)\Tilde{\psi}(r,t)=0
\end{equation}
Where \begin{equation}
    f^{\mu \nu}= \frac{\rho_{\rm 0}}{c_{\rm s0}^2}\begin{bmatrix}
    1 & u_{\rm 0}^1 & u_{\rm 0}^2&u_{\rm 0}^3\\
    u_{\rm 0}^1&u_{\rm 0}^1 u_{\rm 0}^1-{c_{\rm s0}}^2&u_{\rm 0}^1 u_{\rm 0}^2 &u_{\rm 0}^1u_{\rm 0}^3\\
    u_{\rm 0}^2&u_{\rm 0}^2 u_{\rm 0}^1& u_{\rm 0}^2u_{\rm 0}^2-{c_{\rm s0}}^2 &u_{\rm 0}^2u_{\rm 0}^3\\
    u_{\rm 0}^3&u_{\rm 0}^3 u_{\rm 0}^1&u_{\rm 0}^3 u_{\rm 0}^2 &u_{\rm 0}^3u_{\rm 0}^3-{c_{\rm s0}}^2\\
    
    \end{bmatrix}
\end{equation}
where $(u_{\rm 0}^1,u_{\rm 0}^2,u_{\rm 0}^3)$ are the components of the 3-velocity $\mathbf{u_{\rm 0}}$ ($\mathbf{u_{\rm 0}}=-\nabla\psi_{\rm 0}$).\\
For the case of spherically symmetric Bondi flow, the only  available velocity component is the radial velocity component  which we denote simply with $u$. The continuity equation for this spherically symmetric flow is given by 
\begin{equation}
 \frac{\partial \rho}{\partial t}+\frac{1}{r^2}\frac{\partial}{\partial r} \left( \rho u r^2 \right)=0
\end{equation}
For steady state we get, $\rho u r^2 = {\rm constant}$, which essentially is the mass accretion rate as was obtained through eq. (\ref{mass accretion rate})\\
In this scenario we consider a perturbation of the mass accretion rate of the fluid \begin{equation}
  \Sigma(r,t)=\Sigma_{\rm 0}(r)+\epsilon \Tilde{\Sigma}(r,t)
\end{equation}
The equation of perturbation $\Tilde{\Sigma}(r,t)$ is of the form
\begin{equation}
\label{eq:wave}
    \partial_\mu(\Sigma^{\mu \nu}\partial_\nu)\Tilde{\Sigma}(r,t)=0
\end{equation}
where \begin{equation}
    \Sigma^{\mu \nu}= \frac{u_{\rm 0}}{\Sigma_{\rm 0}}\begin{bmatrix}
    1 & u_{\rm 0}\\
    u_{\rm 0}&{u_{\rm 0}}^2-c_{\rm s0}^2\\
    \end{bmatrix}
\end{equation}

\section{Analogue Spacetime}
The equation of propagation of a massless scalar field $(\Phi)$ in a curved spacetime takes the form:
\begin{equation}
     \partial_\mu(\sqrt{-g}g^{\mu \nu}\partial_\nu \Phi )=0
\end{equation}
we observe that this equation looks similar to eq. (\ref{eq:pertub}) if we identify 
\begin{equation}
\label{eq:can_compact}
    f^{\mu \nu}=\sqrt{-g}g^{\mu \nu}
\end{equation}where $g=\det(g_{\mu \nu})$.
A direct comparison and some algebraic manipulation lead to an expression for the space-time
metric governing the propagation of the linear perturbations embedded within the fluid
\begin{equation}
     g_{\mu \nu}=- \frac{\rho_{\rm 0}}{c_{\rm s0} }\begin{bmatrix}
    -({c_{\rm s0}}^2-{\mathbf{u}_{\rm 0}^2}) & -u_{\rm 0}^1 & -u_{\rm 0}^2&-u_{\rm 0}^3\\
    -u_{\rm 0}^1&1&0 &0\\
    -u_{\rm 0}^2&0& 1 &0\\
    -u_{\rm 0}^3&0&0 &1\\
    
    \end{bmatrix}
\end{equation}
The corresponding line element is given by:
% \begin{equation}
% \label{eq:anale}
%     {ds}^2=g_{\mu\nu}dx^\mu dx^\nu=-\frac{\rho_{\rm 0}}{c_{\rm s0}}[-{c_{\rm s0}}^2dt^2+(dx^i-u_{\rm 0} ^i dt )\delta_{ij}(dx^j-u_{\rm 0}^j dt)]
% \end{equation} 
\begin{align}
\label{eq:anale}
ds^2 & = g_{\mu\nu}dx^\mu dx^\nu \notag \\
&= -\frac{\rho_{\rm 0}}{c_{\rm s0}} \bigg[
    -{c_{\rm s0}}^2 dt^2 
    + (dx^i - u_{\rm 0}^i dt) \delta_{ij} (dx^j - u_{\rm 0}^j dt)
\bigg]
\end{align}
This line element is similar to the Schwarzschild line element in Painlevé-Gullstrand coordinates and has a nonzero
curvature. However, since eq. (\ref{eq:pertub}) holds for the acoustic perturbations, the curved spacetime is visible only to them and the background spacetime remains Newtonian. The emergence of such an effective metric in the
context of an inviscid, irrotational, classical fluid (known as the ``acoustic metric") has been recognized since Unruh’s
pioneering work in 1981 \cite{unruh1981experimental}, and has further been formalized by Visser \cite{visser1998acoustic}. Models using this concept are called
`analogue gravity’ models (for a detailed review see \cite{visser_review}).  In astrophysical accretion, analogue gravity models are
unique (see \cite{tapas_review}), as they may involve both a gravitational event horizon and an acoustic horizon (if the accretor is a black
hole). Once the analogy is established for a given configuration, certain methodologies as developed in the theory of
General Relativity can be used to study the causal structure of the spacetime. Earlier, a similarity between the sound
horizon and a black hole event horizon was noted. It is particularly interesting to establish that the sound horizon
is indeed a null horizon in the causal structure of the acoustic spacetime. With a steady-state solution (and a phase
portrait), we can calculate the acoustic spacetime at every $x$. \subsection{Carter-Penrose diagram}
Carter-Penrose diagrams (CP hereafter) is a standard tool of visualizing the causal structure of the entire spacetime and is extensively used in GR and differential geometry. The basic idea is to use conformal transformation and compactification to obtain a finite representation of the infinite spacetime. A detailed account of CP in the context of GR can be found in \cite{townsend1997black, fre2013gravity, hawking2023large} . In the context of analogue spacetime, CP was first used in \cite{cp_visser} . A detailed description of how such diagrams can be drawn and their use in the context of the analogue gravity context can be found in \cite{cp_visser, maity2022carter} . In this section we only present a brief description of the basic methodology in connection to how one can use CP to identify sonic horizons.
\begin{figure*}[ht]
    \centering
    \includegraphics[width=0.8\linewidth]{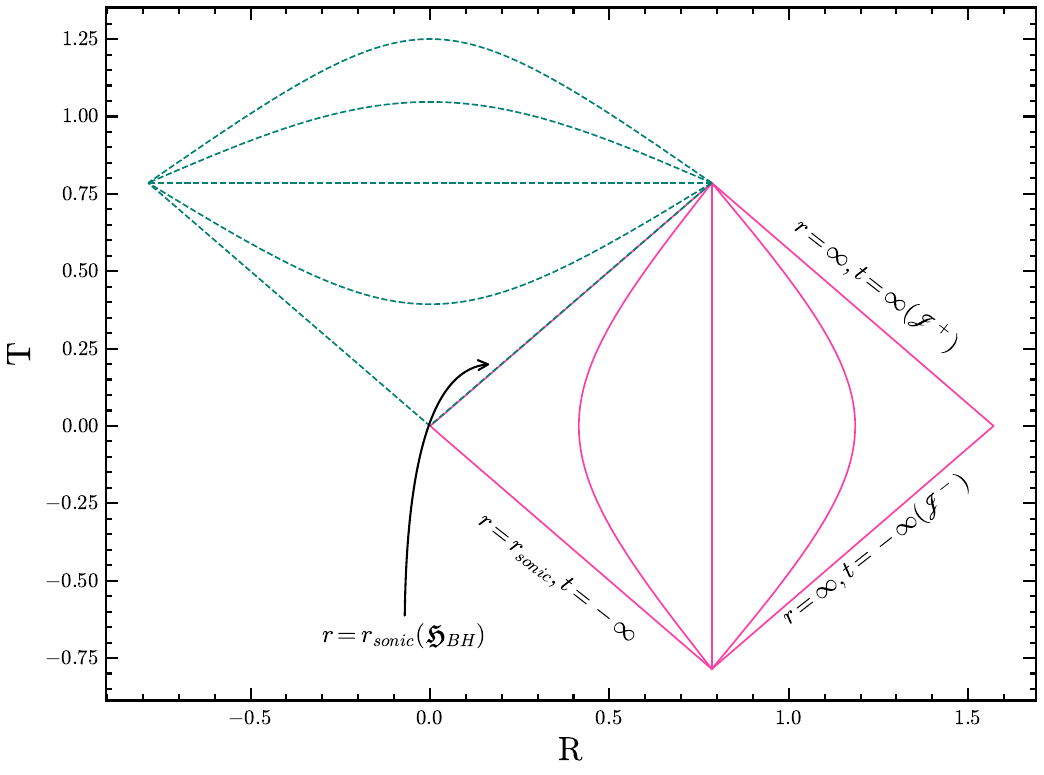}
    \caption{Carter-Penrose Diagram for analogue spacetime obtained from the stationary accretion solution with PW potential (fig. (\ref{fig:a})). The coordinate $T$ and $R$ are defined through eq. (\ref{treqs})- (\ref{treqs2})}
    \label{fig:cpd}
\end{figure*}\\
One can write down the line element of analogue spacetime using  null coordinates (soundlike in case of analogue space-time) $ds^2 =0$, which yields
\begin{equation}
\left(dt-A_{+} (r) dr\right)\left(dt-A_{-} (r) dr\right)=0
\end{equation}
where,
\begin{equation}
A_{\pm} = \frac{-g_{tr} \pm \sqrt{g_{tr}^2-g_{rr}g_{tt}}}{g_{tt}} .
\end{equation}
Thus, instead of using the usual $(t,r)$ co-ordinates, we use new  null co-ordinates $(\chi ,\omega)$, such that
\begin{eqnarray}
\label{null_coordinates_differential1}
d\omega = dt -A_{+}(r)dr\\
\label{null_coordinates_differential2}
d\chi = dt -A_{-}(r)dr
\end{eqnarray}
Using the coordinate transformations introduced in (\ref{null_coordinates_differential1}) and (\ref{null_coordinates_differential2}), the line element (\ref{eq:anale}) assumes the form (and for the present case, we can move to a spherical polar coordinate basis):
\begin{equation}
ds^2 = g_{tt} d\chi d\omega ,
\end{equation}
by expanding $A_- (r)$ and $A_+ (r)$ up to first order of $(r-r_{\rm c} )$. One can expand $u_{\rm 0}$ near $r_{\rm c}$ to obtain:
\begin{equation}\label{expansion}
u_{\rm 0} (r) = -u_{\rm 0 c}  + \left| \frac{du}{dr}\right|_{r_{\rm c} }(r-r_{\rm c} ) + O\left( (r-r_{\rm c})^2\right),
\end{equation}
where the negative sign of the $u_c$ implies the flow is towards the accretor. Similarly, expanding $c_{s_{\rm 0}}$, one obtains:
\begin{equation}\label{divergent_factor_upto_first_order}
u_{\rm 0}^2- c_{\rm s0}^2 \approx -2 (u_{\rm 0c})\left( u_{\rm 0}^{\prime} - c_{\rm s0}^{\prime}\right) (r-r_{\rm c} ),
\end{equation}
considering only the leading order terms, where $\prime$ (prime) denotes derivative with respect to $r$.\\
One needs to expand $A_- (r)$ and $A_+ (r)$ up to linear order of $(r-r_{\rm c} ) $. Note that $g_{tt} \propto (u_{\rm 0}^2 - c_{\rm{s}0}^2)$ is very small near $r_{\rm c}$, implying $|g_{tt}g_{rr}/g_{tr}^2| \ll 1$. Thus,
\begin{eqnarray}
A_+ (r) & = & \frac{ -g_{tr} + g_{tr}\left( 1- \frac{g_{tt}g_{rr}}{g_{tr}^2}\right)^{1/2}}{g_{tt} } \\
& \approx & -\frac{g_{rr}}{2g_{tr}}
\end{eqnarray}
and
\begin{eqnarray}
A_- (r) & = & \frac{ -g_{tr} - g_{tr}\left( 1- \frac{g_{tt}g_{rr}}{g_{tr}^2}\right)^{1/2}}{g_{tt} } \\
& \approx & -\frac{2g_{tr}}{g_{tt}}\\
& = & \frac{1}{\varkappa} \left(  \frac{1}{r-r_{\rm c}} \right)
\end{eqnarray}
where,
\begin{equation}
\varkappa = (u_{\rm 0}^{\prime})\vert_{r_{\rm c}} - (c_{\rm s0}^{\prime})\vert_{r_{\rm c}}.
\end{equation}
Although
\begin{equation}
\chi \approx t- \frac{1}{\varkappa}\ln|r-r_{\rm c} | 
\end{equation}
shows a logarithmic divergence at $r = r_{\rm c}$, the forms of $g_{rr}$ and $g_{tr}$ ensure that
\begin{equation}
\omega = t + \int \frac{g_{rr}}{2g_{tr}} dr
\end{equation}
is still regular.
Near critical points one thus obtains,
\begin{equation}\label{divergent_exponent}
e^{-\varkappa \chi} \propto e^{-\varkappa t}\left\vert r-r_{\rm c} \right\vert \propto  e^{-\varkappa t} (u_{\rm 0}^2-c_{\rm s0}^2)
\end{equation}
One can compare the acoustic null coordinates with those of the Schwarzschild space time, and can deduce a coordinate removing the singularity of the metric element at the critical point. The transformation equations are:
\begin{eqnarray}
\left.\begin{aligned}
U(\chi) &= -e^{-\varkappa\chi}\\
W(\omega) &= e^{\varkappa\omega}
\end{aligned}\right.
\end{eqnarray}
Finally, one can compactify the infinite space into a finite patch, using new coordinate \((T, R)\), such that:
\begin{eqnarray}
\label{treqs}
T= \tan^{-1} (W)+\tan^{-1} (U)\\
\label{treqs2}
R= \tan^{-1} (W)-\tan^{-1} (U).
\end{eqnarray}
In $(T,R)$ coordinates, lines where $r=\rm{constant}$ are drawn which allow the resulting diagram to represent the causal structure of the original spacetime in a compactified region. The resulting diagram, as shown in Fig. \ref{fig:cpd}, is known as the Carter-Penrose diagram. In obtaining fig. (\ref{fig:cpd}) we have used the stationary solutions obtained for the PW potential. However, the causal structure for the other potentials as mentioned here are similar, since they all lead to the same analogue metric.\\
A certain lemma in differential geometry is useful to interpret fig. (\ref{fig:cpd}).
\textit{If two matrices $G$ and $g$ on the same manifold $\mathscr{M}$ are conformally related, then the null geodesics with respect to metric $G$ are null geodesics also with respect to the metric $g$ and vice-versa.}
Thus any perturbation propagating with the sound-speed is null-like in the acoustic space-time. \\ One can define the boundary of the mapping $\psi$ of the entire analogue space-time $\mathscr{M}$ as
\begin{equation}
\partial \psi\left( \mathscr{M} \right) = i^0 \bigcup \mathscr{J}^+ \bigcup \mathscr{J}^-
\end{equation}
where
\begin{enumerate}
\item $i_{\rm 0}$, the \textit{Spatial Infinity}, is the endpoint of the $\psi$ image of all space-like curves in ($\mathscr{M}$, $g$).
   
\item $\mathscr{J}^+$, formally known as the \textit{Future Causal Infinity} is the endpoint of the $\psi$ image of all future directed causal curves in ($\mathscr{M}$, $g$).
   
\item $\mathscr{J}^-$, \textit{Past Causal Infinity}, is the endpoint of the $\psi$ image of all past directed causal curves in ($\mathscr{M}$, $g$).
\end{enumerate}
The sound horizon can be readily identified as a null hypersurface from the fig. (\ref{fig:cpd}). Evidently, no acoustic perturbation, created within the sound horizon (green, dashed region in the fig. (\ref{fig:cpd}) can cross the sound horizon and escape to a large distance away from the accretor (pink, solid region in the fig. (\ref{fig:cpd}). The sound horizon is a null horizon in the acoustic space-time. Hence, the sound horizon behaves like an event horizon for acoustic perturbations.
\section{Analogue surface gravity}
\noindent
In the context of Einstein's general relativity, the term ``Surface gravity" actually refers to the gravitational pull on a fluid element upon event horizon (or Killing horizon) around the black hole, as perceived by a distant observer at infinity. From the observer reference frame, surface gravity is the specific force needed to keep the fluid element on the horizon at rest. For our reference to the analogue space-time of black hole, ``Acoustic surface gravity" is the analogous specific force experienced by the sound waves approaching the acoustic horizon beyond which it can't escape. Hence the redshift of the sound waves near horizon depends on this acoustic surface gravity. In the following, we will write the expression of acoustic surface gravity in acoustic space-time which mimics the Schwarzschild metric, and later will use that expression for the above mentioned four pseudo-Newtonian gravitational potentials to see the graphical nature of the quantities.  
The expression of general relativistic energy-momentum  tensor $T^{\mu \nu}$ for any ideal fluid is given by,
\begin{equation}
    T^{\mu \nu} = \left[ (\epsilon + p)u^{\mu} u^{\nu} + pg^{\mu \nu} \right]
    \label{E_M_exp}
\end{equation}

where, $\epsilon$ is the energy density of fluid, $p$ is the pressure, and $g^{\mu \nu}$ is the space-time metric.

The four divergence of the tensor $T^{\mu \nu}$ gives us Euler equation,
\begin{equation}
    T^{\mu \nu}_{;\nu} = 0
    \label{Euler_eq}
\end{equation}

Here we will be dealing with our acoustic metric, hence there is no time dependence. So we can choose our Killing vector as,

\begin{equation}
    \psi^{\mu} = \delta^{\mu}_{t}
    \label{killing1}
\end{equation}

The acoustic surface gravity expression has been derived in \cite{TapasAbraham:2005ah} (refer to sec.$6$ therein) and is given by,
\begin{equation}
    \kappa = \left \vert\sqrt{\frac{\psi^{\mu} \psi_{\mu}}{-g_{rr}}} \frac{1}{1 - c_{s}^2} \left(\frac{du}{dr} - \frac{dc_{s}}{dr} \right)\right \vert_{r_{\rm c}}
    \label{kappa_main}
\end{equation}

We know, 
$$\sqrt{\psi^{\mu} \psi_{\mu}} = \sqrt{G_{tt}} = \sqrt{(1 + 2\Phi)} \approx (1 + \Phi)$$
 and,
 $$g_{rr} = -1$$

\begin{figure*}[ht]
    \centering
    \includegraphics[width=0.8\linewidth]{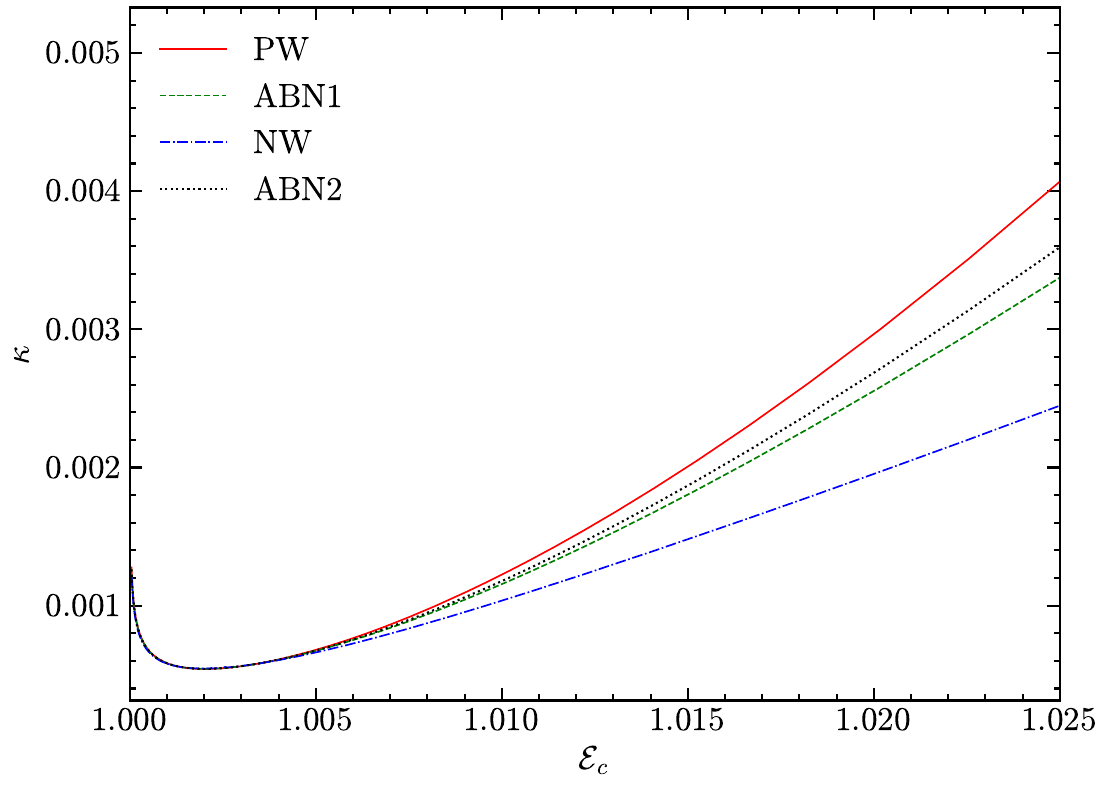}
    \caption{Variation of surface gravity (defined through eq. (\ref{kappa_final})) with energy $\mathcal{E}_{c} (=\mathcal{E})$ for a fixed $\xi  = 1$ and for different pseudo-potentials defined through eqs (\ref{eq:pw})-(\ref{eq:abn2})}
    \label{fig:kappa_pot}
\end{figure*}

\begin{figure*}[ht]
    \centering
    \includegraphics[width=0.8\linewidth]{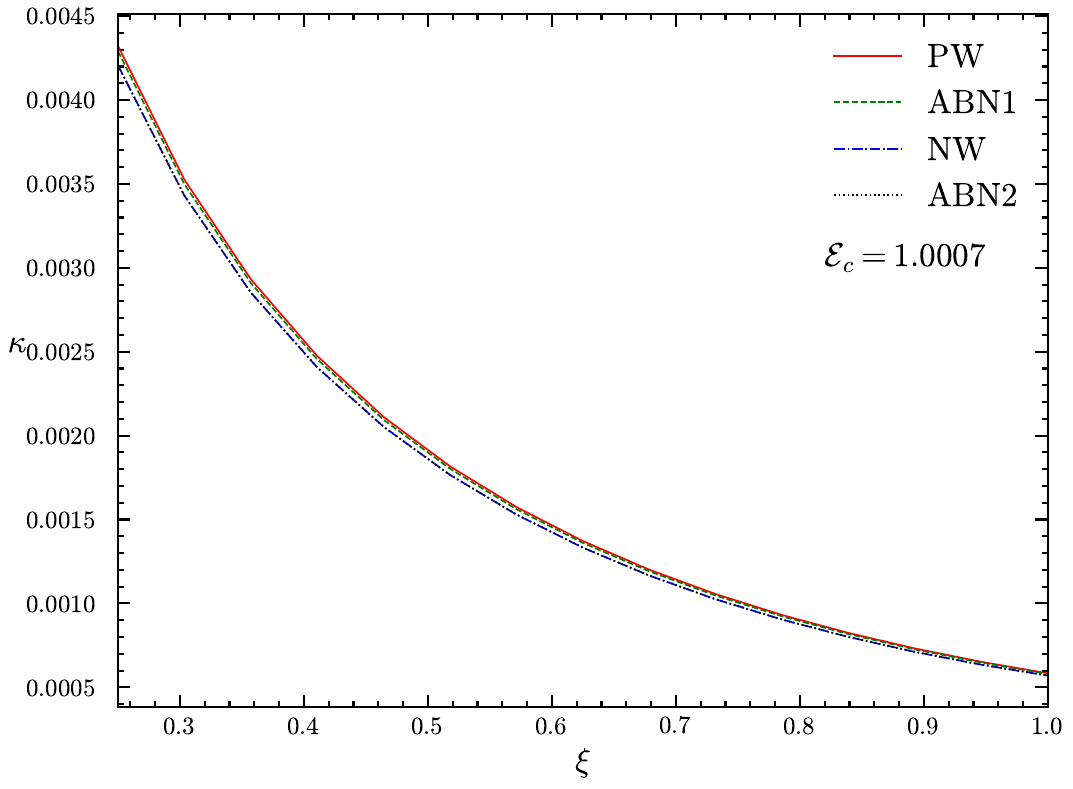}
    \caption{Variation of surface gravity ($\kappa$) with composition (proportion of different components as determined by the different values $\xi$ as defined in sec. \ref{sec:meos}) of fluid for a demonstrative $\mathcal{E}_c (=\mathcal{E})$. Refer to eq. (\ref{kappa_final}) for the definition of surface gravity. The relevant pseudo-potentials are defined through eqs (\ref{eq:pw})-(\ref{eq:abn2}).}
    \label{fig:kappa_xi}
\end{figure*}

Hence the final expression for the acoustic surface gravity at the horizon $r_{c}$ is (see section $3$ of \cite{Bilic:2012yh} for details):

\begin{equation}
    \kappa = \left \vert\sqrt{(1 + \phi)} \frac{1}{(1 - c_{s}^2)} \left(\frac{du}{dr} - \frac{dc_{s}}{dr} \right)\right \vert_{r_{\rm c}}
    \label{kappa_final}
\end{equation}
Variation of surface gravity ($\kappa$) with the energy ($\mathcal{E}_c$) is shown in fig. (\ref{fig:kappa_pot}). We also checked the variation of surface gravity with the composition of the flow for a particular energy which is shown in fig. (\ref{fig:kappa_xi}).

% \begin{figure*}
%     \centering
%     \begin{tabular}{cc}
%         \subcaptionbox{Caption 1\label{fig:4a}}{\includegraphics[width=0.45\textwidth]{image1.png}} &
%         \subcaptionbox{Caption 2\label{fig:4b}}{\includegraphics[width=0.45\textwidth]{image2.png}} \\
%         \subcaptionbox{Caption 3\label{fig:4c}}{\includegraphics[width=0.45\textwidth]{image3.png}} &
%         \subcaptionbox{Caption 4\label{fig:4d}}{\includegraphics[width=0.45\textwidth]{image4.png}} \\
%     \end{tabular}
%     \caption{Variation of $\kappa$ with $\mathcal{E}$ for four different potentials}
%     \label{fig:main}
% \end{figure*}

\section{Standing and traveling wave analysis}
Consider a trial wave function as a solution to the wave equation (eq.(\ref{eq:wave})) for the Bondi flow:
\begin{equation}
    \Tilde{f}=f_\omega(r)\exp{(-i\omega t)}
\end{equation} The amplitude part $g_\omega(r)$ satisfies the equation: \begin{align}
\label{eq:omega1}
    -\omega^2f^{tt}&f_\omega-i\omega(f^{tr}\partial_rf_\omega+f_\omega\partial_rf^{rt}+f^{rt}\partial_rf_\omega) \notag \\ 
    &+(\partial_rf_\omega\partial_rf^{rr}+f^{rr}\partial_{rr}f_\omega)=0
\end{align} 
where $f^{tt}=\frac{u_{\rm 0}}{f_{\rm 0}}$, $f^{rt}=f^{tr}=\frac{{u_{\rm 0}}^2}{f_{\rm 0}}$ and $f^{rr}=\frac{u_{\rm 0}}{f_{\rm 0}}({u_{\rm 0}}^2-{c_{\rm s0}}^2)$. This solution can represent a standing wave as well as a traveling wave.\subsection{Standing Wave Analysis}
If the accretor is an astrophysical object with a well defined physical boundary, the perturbations should vanish at the two extremes, i.e., far away from the accretor and on its surface. A standing wave solution is useful in such scenarios. As a transonic flow becomes supersonic at the sound horizon, the only way it can become subsonic again is through a discontinuous shock transition. But a standing wave solution is continuous. So, for the flow to have a standing wave solution it has to be subsonic throughout. \newline
Substituting the values of the metric element $f^{\mu \nu}$ in eq. (\ref{eq:omega1}) we get, 
%\begin{align}\noindent
%&-\omega^2u_{\rm 0}f_\omega-2i\omega\left({u_{\rm 0}}^2\partial_rf_\omega \right. \notag \\ &+\left.u_{\rm 0}f_\omega\partial_ru_{\rm 0}\right)+\partial_r(u_{\rm 0}({u_{\rm 0}}^2-{c_{\rm s0}}^2)\partial_rf_\omega)=0. \label{eq:89}
%\end{align}
%Which leads to:
\begin{equation}\noindent
-\omega^2f_\omega-2i\omega\partial_r(u_{\rm 0}f_\omega)
+\frac{1}{u_{\rm 0}}\partial_r\left(u_{\rm 0}({u_{\rm 0}}^2-{c_{\rm s0}}^2)\partial_rf_\omega\right)=0 \label{sw_1}
\end{equation}
This is a quadratic equation in $\omega$, which can be solved.
Let the two boundaries be $r_1$ and $r_2(> r_1)$ such that $f_\omega(r_1) = f_\omega(r_2) = 0$. Now multiplying eq. (\ref{sw_1}) with $u_{\rm 0}f_\omega$  and integrating in the range $r_1 < r < r_2$ we get
\begin{align}
&-\omega^2\int u_{\rm 0}f_\omega^2 dr-iw\int  \partial_r[{u_{\rm 0}}^2{f_\omega}^2]dr \notag \\
&+\int f_\omega\partial_r[u_{\rm 0}({u_{\rm 0}}^2-{c_{\rm s0}}^2)\partial_rf_\omega]dr=0\\
&\omega^2\int u_{\rm 0}f_\omega^2 dr+\int [u_{\rm 0}({u_{\rm 0}}^2-c_{\rm s0}^2)({\partial_rf_\omega})^2]dr=0\\
&\omega^2=-\frac{\int [u_{\rm 0}({u_{\rm 0}}^2-c_{\rm s0}^2)({\partial_rf_\omega})^2]dr}{\int u_{\rm 0}f_\omega^2 dr}
\end{align}

As the flow is subsonic everywhere, $u_{\rm 0} < c_{\rm s0}$, $\omega$ must be real in this region and we get oscillatory solution for standing wave with any diverging term.
\subsection{Traveling Wave Analysis}
If the accretor happens to be a black-hole, there is no hard surface and any flow that reaches the event horizon is supersonic. In such a scenario the analysis described in the previous section is not valid. The wave in this scenario is a traveling one. The wavelength of such a wave is small compared to the radius of the accretor.
Eq. (\ref{eq:omega1}) then takes the form: 
\begin{align}
\label{eq:omegae}
    ({u_{\rm 0}}^2&-{c_{\rm s0}}^2){f_\omega}''\notag \\
    &+[3u_{\rm 0}\partial_ru_{\rm 0}-\frac{1}{u_{\rm 0}}\partial_r(u_{\rm 0}c_{\rm s0}^2)-2i\omega u_{\rm 0}]f_\omega'\notag \\
    &-(2i\omega\partial_ru_{\rm 0}+\omega^2)f_\omega = 0
\end{align}
Employing the WKB approximation method we can approximate the solution in  a power series of $\omega$:
\begin{equation}
\label{eq:fomega}
    f_\omega(r)=\exp \left[ \sum_{n=-1}^\infty \frac{k_n(r)}{\omega^n}\right]
\end{equation}
Substituting this value in eq. (\ref{eq:omegae}) and collecting the coefficients of $\omega^0$, $\omega^1$ and $\omega^2$ we get the following equations:
\begin{equation}
      ({u_{\rm 0}}^2-c_{\rm s0}^2){\left(\frac{dk_{-1}}{dr}\right)}^2  -2iu_{\rm 0}\frac{dk_{-1}}{dr}-1=0
\end{equation}
\begin{align}
&({u_{\rm 0}}^2-c_{\rm s0}^2)\left(\frac{d^2k_{-1}}{{dr}^2}+2\frac{dk_{-1}}{dr}\frac{dk_{\rm 0}}{dr}\right) \notag\\
&+\left[ 3u_{\rm 0}-\frac{1}{u_{\rm 0}}\partial_r(u_{\rm 0}c_{\rm s0}^2)\right]\frac{dk_{-1}}{dr}-2i\left( u_{\rm 0}\frac{dk_{\rm 0}}{dr}+\partial_ru_{\rm 0}\right)=0
\end{align}
\begin{align}
    &({u_{\rm 0}}^2-{c_{\rm s0}}^2)\left( \frac{d^2k_{0}}{{dr}^2}+2\frac{dk_{-1}}{dr}\frac{dk_1}{dr}+{\left(\frac{dk_{\rm 0}}{dr}\right)}^2\right)\notag \\&+\left[ 3u_{\rm 0}-\frac{1}{u_{\rm 0}}\partial_r(u_{\rm 0}{c_{\rm s0}}^2)\right]\frac{dk_{0}}{dr}-2iu_{\rm 0}\frac{dk_1}{dr}=0
\end{align}
solving for $\frac{dk_{-1}}{dr}$, $\frac{dk_{0}}{dr}$ and $\frac{dk_{1}}{dr}$ and then by integration we get the values as
\begin{equation}
    \begin{split}
        \frac{dk_{-1}}{dr}=\frac{i}{u_{\rm 0}\mp c_{\rm s0}}\\
        k_{-1}=i\int\frac{dr}{u_{\rm 0} \mp c_{\rm s0}}
    \end{split}
\end{equation}
\begin{equation}
    k_{\rm 0}=-\frac{1}{2}\ln{\left(\frac{u_{\rm 0} c_{\rm s0}}{f_{\rm 0}}\right)}
\end{equation}
%When $r \to \infty$, $u_{\rm 0}\mp a_{\rm 0} \to a_\infty$ and $ $then $k_{-1}\sim r$
Thus for $\omega \gg 1$ the first three terms of $f_\omega(r)$ in the expression (eq. (\ref{eq:fomega})) are  $\omega r$, $\ln{r}$ and $1/(\omega r)$ respectively. For large values of $r$:
\begin{equation}
    \omega r \ll \ln{r} \ll \frac{1}{\omega r}
\end{equation}
which leads to:
\begin{equation}
    \omega _{|k_{-1}|}\ll |k_{\rm 0}| \ll \frac{1}{\omega_{|k_1|}}
\end{equation}
Hence, power series in $f_\omega(r)$ does not diverge even as $n$ increases, that is,
\begin{equation}
    \omega^{-n}|k_n(r)| \ll \omega^{-(n+1)}|k_{n+1}(r)|
\end{equation}
and one obtains a traveling wave solution with finite amplitude. Thus, in both the cases of standing and traveling waves, the stability of the steady state solution is ensured.
\section{Conclusion}
In the present work, we find that only one black hole type acoustic horizon forms in the system. This is a consequence of the spherical symmetry. For axially symmetric accretion (accretion disc)  containing small amount of intrinsic flow angular momentum, it has been observed that more than one sonic points may be found in the system, and the accretion flow can undergo sonic state transition thrice; twice from the subsonic to supersonic state via  smooth transitions through saddle type sonic points, and once from supersonic to subsonic via discontinuous transition through a Rankine-Huguenot (\cite{clarke2007principles, landau1987fluid}) type stationary shock. Such a shock location can be identified with an acoustic white hole. In our next series of works, we plan to investigate the analogue metric for axially symmetric accretion flow under the influence of various post Newtonian pseudo Schwarzschild and pseudo Kerr black hole potentials for accretion governed by the equation of state as described in the current paper, and shall investigate how the analogue space time metric can be influenced by the actual black hole metric, by studying the effect of black hole spin angular momentum (the Kerr parameter, to be more specific) in determining the location of the corresponding acoustic horizons and the value of the acoustic surface gravity, respectively. 
\section*{Acknowledgments}
Tuhin Paul acknowledges Harish-Chandra Research Institute for supporting a visit through Research Apex Sub project Astro, during which a part of the work was done.
\appendix
\section{}
\label{appena}
\renewcommand{\theequation}{A\arabic{equation}}
\setcounter{equation}{0}  % Reset equation counter
Using eq. (\ref{eq:pw}) in eq. (\ref{velocity_grad_general}):
\begin{equation}
    \label{eq:dvdr}
    \frac{du}{dr}=\frac{\frac{2c_{\rm s}^2}{r}-\frac{1}{2(r-1)^2}}{u-\frac{c_{\rm s}^2}{u}}=\frac{\mathcal{N}}{\mathcal{D}}\; (\rm say).
\end{equation}
Using eq. (\ref{eq:pw}) and eq. (\ref{eq:dvdr}) in eq. (\ref{eq:dthetadr_gen}) one obtains:
 \begin{equation}
    \label{eq:dthetadr}
    \frac{d\Theta}{dr}=\frac{\frac{2\Theta }{N}\left(\frac{u^2}{r}-\frac{1}{4(r-1)^2}\right)}{c_{\rm s}^2-u^2}
\end{equation}
Similarly, from eq. (\ref{eq:abn1}):
\begin{equation}
    \frac{d\phi_2}{dr} = \frac{1}{2r^2} \left( 1 - \frac{1}{r} \right)^{-1/2}
    \label{phi2_1}
\end{equation}
Putting this value in eq: (\ref{velocity_grad_general}) we get the velocity gradient for the case of ABN1 potential in the form :
\begin{equation}
     \frac{du}{dr} = \frac{ \frac{2c_{\rm s}^2}{r} - \frac{1}{2r^2} \left( 1 - \frac{1}{r} \right)^{-1/2} }{u - \frac{c_{\rm s}^2}{u}}
\end{equation}
Also substituting this value of eq: (\ref{phi2_1}) in eq: (\ref{eq:dthetadr_gen}), we get the expression for $\frac{d\Theta}{dr}$:
\begin{equation}
  \frac{d\Theta}{dr}=\frac{\frac{2\Theta }{N}\left(\frac{u^2}{r}-\frac{1}{4r^2} \left( 1 - \frac{1}{r} \right)^{-1/2}\right)}{c_{\rm s}^2-u^2}
\end{equation}
%\begin{equation}
 %   \frac{d^2\phi_2}{dr^2} = -\frac{1}{r^3}\left( 1 - \frac{1}{r} \right)^{-1/2} - \frac{1}{4r^2}\left( 1 - \frac{1}{r} \right)^{-3/2}
  %  \label{phi2_2}
%\end{equation}

From eq: (\ref{eq:nw}):
\begin{equation}
    \frac{d\phi_3}{dr} = \frac{1}{2r^2} - \frac{3}{2r^3} + \frac{1}{2r^4}
    \label{phi3_1}
\end{equation}
In a similar manner as discussed in the previous case, using eq: (\ref{velocity_grad_general}) the velocity gradient for the case of  NW potential is:
\begin{equation}
     \frac{du}{dr} = \frac{ \frac{2c_{\rm s}^2}{r} - \frac{1}{2r^2} + \frac{3}{2r^3} - \frac{1}{2r^4} }{u - \frac{c_{\rm s}^2}{u}}
\end{equation}
The expression of $\frac{d\Theta}{dr}$ in a similar manner using eq: (\ref{eq:dthetadr_gen}) and eq: (\ref{phi3_1}), 
\begin{equation}
    \frac{d\Theta}{dr}=\frac{\frac{2\Theta }{N}\left(\frac{u^2}{r}- \frac{1}{4r^2} + \frac{3}{4r^3} - \frac{1}{4r^4}
    \right)}{c_{\rm s}^2-u^2}
\end{equation}

%\begin{equation}
 %    \frac{d^2\phi_3}{dr^2} = -\frac{1}{r^3} + \frac{9}{2r^4} - \frac{2}{r^5}
  %   \label{phi3_2}
%\end{equation}

From eq (\ref{eq:abn2}):

\begin{equation}
    \frac{d\phi_4}{dr} = \frac{1}{2r^2} \left(1 - \frac{1}{r} \right)^{-1}
    \label{phi4_1}
\end{equation}
The velocity gradient for the case of ABN2 potential:
\begin{equation}
     \frac{du}{dr} = \frac{ \frac{2c_{\rm s}^2}{r} - \frac{1}{2r^2} \left(1 - \frac{1}{r} \right)^{-1} }{u - \frac{c_{\rm s}^2}{u}}
\end{equation}
The expression of $\frac{d\Theta}{dr}$:
\begin{equation}
     \frac{d\Theta}{dr}=\frac{\frac{2\Theta }{N}\left(\frac{u^2}{r}-\frac{1}{4r^2} \left(1 - \frac{1}{r} \right)^{-1}\right)}{c_{\rm s}^2-u^2}
\end{equation}

%\begin{equation}
 %   \frac{d^2\phi_4}{dr^2} = \frac{2(1-2r)}{4r^4 \left(1 - \frac{1}{r} \right)^2}
  %  \label{phi4_2}
%\end{equation}
\section{}
\label{appenb}

\renewcommand{\theequation}{B\arabic{equation}}
\setcounter{equation}{0}  % Reset equation counter

Using the familiar dynamical systems approach for a general potential function as discussed in eq: (\ref{velocity_grad_general}), we introduce a  variable $\tau$ and write the equation of velocity gradient as:
\begin{equation}
     \frac{du}{dr} = \frac{du/d\tau}{dr/d\tau}=\frac{ \frac{2c_{\rm s}^2}{r} - \frac{d\phi}{dr}}{u - \frac{c_{\rm s}^2}{u}} = \frac{\mathcal{N}}{\mathcal{D}}
   \end{equation}
Then following our usual calculation scheme,
\begin{align}
\label{eq:dyn1_gen}
    \frac{d}{d\tau}(\delta u)&=\frac{\partial \mathcal{N}}{\partial r} \delta r+\frac{\partial \mathcal{N}}{\partial \Theta} \delta \Theta \notag \\ &= \left(-\frac{2c_{\rm s}^2}{r^2}-\frac{d^2\phi}{dr^2}\right)\delta r 
    +\frac{4c_{\rm s}c_{\rm s}'}{r}\delta \Theta,
\end{align}
\begin{align}
\label{eq:dyn2_gen}
     \frac{d}{d\tau}(\delta r) &=\frac{\partial \mathcal{D}}{\partial u} \delta u+\frac{\partial \mathcal{D}}{\partial \Theta} \delta \Theta \notag \\
     &=\left( 1+\frac{c_{\rm s}^2}{u^2}\right)\delta u+\left( - \frac{2c_{\rm s}c_{\rm s}'}{u}\right)\delta \Theta
\end{align}
We use the expression of $\delta \Theta$ from eq. (\ref{delta_theta}):
\begin{equation}
    \delta \Theta=-\frac{\Theta}{Nu}\delta u -\frac{2\Theta}{Nr}\delta r
\end{equation}
Substituting this value in eq. (\ref{eq:dyn1_gen}) and eq. (\ref{eq:dyn2_gen}) we have,
\begin{align}
    \frac{d}{d\tau}(\delta u)&=\left( -\frac{4c_{\rm s}c_{\rm s}'\Theta}{Nur}\right) \delta u \notag \\
    &+ \left(-\frac{2c_{\rm s}^2}{r^2}-\frac{d^2\phi}{dr^2}-\frac{8c_{\rm s}c_{\rm s}'\Theta}{Nr^2}\right)\delta r
\end{align}
\begin{equation}
     \frac{d}{d\tau}(\delta r)=\left( 1+\frac{c_{\rm s}^2}{u^2}+\frac{2c_{\rm s}c_{\rm s}'\Theta}{Nu^2}\right)\delta u+\frac{4c_{\rm s}c_{\rm s}'\Theta}{Nur}\delta r
\end{equation}
\begin{align}
    \begin{pmatrix}
         \frac{d(\delta u)}{d \tau} \\
         \frac{d(\delta r)}{d \tau}
     \end{pmatrix} &=
     \begin{pmatrix}
         -\frac{4c_{\rm s}c_{\rm s}'\Theta}{Nur} & -\frac{2c_{\rm s}^2}{r^2}-\frac{d^2\phi}{dr^2}-\frac{8c_{\rm s}c_{\rm s}'\theta}{Nr^2} \\
         1+\frac{c_{\rm s}^2}{u^2}+\frac{2c_{\rm s}c_{\rm s}'\Theta}{Nu^2} & \frac{4c_{\rm s}c_{\rm s}'\Theta}{Nur} 
     \end{pmatrix}\notag \\ & \times\begin{pmatrix}
             \delta u \\
             \delta r
         \end{pmatrix}
\end{align}
Writing this as $Y'=A_0Y$ and assuming the solutions in the form $\delta u \sim \exp{\Omega_0 \tau} $ and $\delta r\sim \exp{\Omega_0 \tau} $, the 
eigenvalue equation for this dynamical system of equations is obtained as, $A_0Y = \Omega_0 Y$. The characteristic equation $\det (A_0 - \Omega_0 I) = 0$ gives the  eigenvalues and results in a quadratic equation
\begin{equation}
    \label{eq:om_quad_gen}
    \Omega_0^2+ \Tr(A_0)\Omega_0+\Delta_0=0
\end{equation}

where $\Delta_0 = \det (A_0)$. 
The nature of the roots of eq. (\ref{eq:om_quad_gen}) depend on the numerical values of $\Delta_0$ and the discriminant $D_0 = ({\rm Tr} A_0)^{2}-  4{\rm \Delta_0}$ and yields the nature of the critical points.\\
In the case of the general potential, the calculation of the matrix elements and the values of ${\rm Tr}(A_0)$ and ${\rm \Delta_0}$, at the critical points $r=r_{\rm c}$, $u=u_{\rm c}$ and $c_{\rm s}=c_{\rm sc}$ reveals the nature of the critical point. We remark, 

\begin{align}
 & \left. {\rm \Tr}(A_0)\right \vert_{\substack{u=u_{\rm c} \\ c_{\rm s}=c_{\rm sc}}}=0 \notag\\
&\left .{\rm \Delta_0} \right \vert_{\substack{u=u_{\rm c} \\ c_{\rm s} = c_{\rm sc}}} =  \left[ \frac{2}{r}\frac{d \phi}{dr}  
+ 2\frac{d^2\phi}{dr^2} \right.\notag \\ & \left.\left.  +\frac{4c_{\rm s}c_{\rm s}'\Theta}{N}\left(\frac{10}{r^2}+\frac{2}{r}\frac{1}{\frac{d\phi}{dr}}\frac{d^2\phi}{dr^2}\right) \right]\right \vert_{\substack{r=r_{\rm c}\\
       u=u_{\rm c}\\
       c_{\rm s}=c_{\rm sc}}}
\end{align}

Arguing in the similar manner as in section \ref{dyn-sec}, we conclude that in order to ensure a saddle type critical point we need to get two real roots of opposite sign and hence we must have $\Delta_0 < 0$.

%\bibliographystyle{apsrev4-2}
%\bibliography{ref}
\end{document}